\documentclass[runningheads]{llncs}
\usepackage[utf8]{inputenc}
\usepackage[T1]{fontenc}

\usepackage{color}
\usepackage{graphicx,caption,subcaption,amsmath,amssymb,tabularx,array,booktabs} %
\usepackage{setspace}
\usepackage{xcolor}
\usepackage{multirow,tabularx}
\usepackage{array}
\newcolumntype{P}[1]{>{\raggedright\arraybackslash}p{#1}}

\usepackage{graphicx}
\usepackage{wrapfig}
\usepackage{lscape}
\usepackage{rotating}

\newcommand{\nop}[1]{}
\usepackage{tablefootnote}

\usepackage{listings}
\usepackage{color}
\usepackage{booktabs}
\usepackage{amssymb}
\usepackage{subcaption}
\usepackage{caption}

\usepackage{booktabs,graphicx}
\usepackage{float}
\usepackage{enumitem}
\usepackage{array}
\newcolumntype{P}[1]{>{\raggedright\arraybackslash}p{#1}}
\usepackage{wrapfig,threeparttable,makecell,listings}

\usepackage{tabularx}
\usepackage{makecell} %
\usepackage{threeparttable,caption,booktabs}

\usepackage{listings}
\usepackage{chngcntr}
\AtBeginDocument{\counterwithout{lstlisting}{section}}

\title{SemOpenAlex: The Scientific Landscape in 26~Billion RDF~Triples}
\titlerunning{SemOpenAlex: The Scientific Landscape in 26~Billion RDF~Triples}

\author{
Michael Färber\inst{1}\textsuperscript{\orcid{0000-0001-5458-8645}} %
\and David Lamprecht\inst{1}\textsuperscript{\orcid{0000-0002-9098-5389}}
\and Johan Krause\inst{1}\textsuperscript{\orcid{0000-0002-5080-3587}}
\and Linn Aung\inst{2}\textsuperscript{\orcid{0009-0009-4338-0983}}
\and Peter~Haase\inst{2}\textsuperscript{\orcid{0000-0002-7561-7000}}
}

\usepackage{todonotes}

\PassOptionsToPackage{hyphens}{url}\usepackage[colorlinks=false, pdfborder={0 0 0}]{hyperref}
\newcommand{\keywords}[1]{\par\addvspace\baselineskip
\noindent\keywordname\enspace\ignorespaces#1}

\newcommand{\orcid}[1]{\href{https://orcid.org/#1}{\includegraphics[width=10pt]{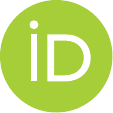}}}

\authorrunning{M. Färber et al.}
\institute{Institute AIFB, Karlsruhe Institute of Technology (KIT), Karlsruhe, Germany\\\email{michael.faerber@kit.edu}, 
\email{david.lamprecht@student.kit.edu}, 
\email{johan.krause@student.kit.edu} %
\and
metaphacts GmbH, Walldorf, Germany\\\email{la@metaphacts.com}, \email{ph@metaphacts.com}
}

\usepackage[none]{hyphenat}
\newcolumntype{P}[1]{>{\centering\arraybackslash}p{#1}}
\newcolumntype{R}[1]{>{\raggedleft\arraybackslash}p{#1}}
\newcolumntype{L}[1]{>{\raggedright\arraybackslash}p{#1}}
\newcolumntype{C}[1]{>{\centering\arraybackslash}p{#1}}

\begin{document}
\maketitle

\begin{abstract}
We present \emph{SemOpenAlex}, an extensive RDF knowledge graph that contains over 26 billion triples about scientific publications and their associated entities, such as authors, institutions, journals, and concepts. SemOpenAlex is licensed under CC0, providing free and open access to the data. We offer the data through multiple channels, including RDF dump files, a SPARQL endpoint, and as a data source in the Linked Open Data cloud, complete with resolvable URIs and links to other data sources. Moreover, we provide embeddings for knowledge graph entities using high-performance computing. SemOpenAlex enables a broad range of use-case scenarios, such as exploratory semantic search via our website, large-scale scientific impact quantification, and other forms of scholarly big data analytics within and across scientific disciplines. Additionally, it enables academic recommender systems, such as recommending collaborators, publications, and venues, including explainability capabilities. Finally, SemOpenAlex can serve for RDF query optimization benchmarks, creating scholarly knowledge-guided language models, and as a hub for semantic scientific publishing. \\[0.2cm]
\begin{tabular}{ll}
     \textbf{Data \& Services: }&\url{https://semopenalex.org}\\ 
                  & \url{https://w3id.org/SemOpenAlex} \\
     \textbf{Code:}&\url{https://github.com/metaphacts/semopenalex/}\\
     \textbf{Data License:}&\href{https://creativecommons.org/publicdomain/zero/1.0/}{Creative Commons Zero (CC0)} \\
      \textbf{Code License:}&\href{https://opensource.org/license/mit/}{MIT License} \\
\end{tabular}
\keywords{Scholarly Data, Open Science, Digital Libraries} %
\end{abstract}

\section{Introduction}
\label{chapter_intro}

With the increasing number of scientific publications, staying up-to-date with current research presents a significant challenge. For instance, in 2022 alone, more than 8 million scientific publications were registered \cite{priem_openalex_2022}. To explore related scholarly entities such as authors and institutions, researchers rely on a range of methods from search interfaces to recommendation systems \cite{auer_towards_2018,christensen_wissenschaftliche_2022}. One effective way to model the underlying scholarly data is to represent it as an RDF knowledge graph (KG). Doing so facilitates standardization, visualization, and interlinking with Linked Data resources \cite{hogan_knowledge_2021_book}. Consequently, scholarly KGs play a pivotal role in transforming document-centric scholarly data into interconnected and machine-actionable knowledge structures \cite{auer_towards_2018}.

However, available scholarly KGs have %
one or several of the following limitations. 
Firstly, they rarely contain an exhaustive catalog of publications across all disciplines \cite{peroni_opencitations_2020}. Secondly, they often cover only certain disciplines, such as computer science \cite{aleman-meza_swetodblp_2007}. Thirdly, they are not regularly updated, rendering many analyses and business models obsolete \cite{farber_makg_2019}. Fourthly, they often contain usage restrictions \cite{waltman_special_2020}. Lastly, even if they fulfill these requirements, they are not available according to W3C standards such as RDF \cite{manghi_new_2021,priem_openalex_2022}. These issues hinder the application of scientific KGs on a broad scale, such as in comprehensive search and recommender systems, or for scientific impact quantification. For instance, the 
Microsoft Academic Graph was discontinued in 2021 \cite{microsoft_research_next_2021}, which hinders further updates to its derivative in RDF, the Microsoft Academic Knowledge Graph (MAKG) \cite{farber_makg_2019}. This leaves a gap that the novel OpenAlex dataset 
aims to fill \cite{priem_openalex_2022}. However, the data in OpenAlex is not available in RDF and does not comply with Linked Data Principles \cite{berners-lee_linked_2006}. Consequently, OpenAlex cannot be considered a KG, which makes semantic queries, integration into existing applications, or linking to additional resources non-trivial. At first glance, integrating scholarly data about scientific papers into Wikidata and thus contributing to the WikiCite initiative may seem like an obvious solution. 
However, apart from the dedicated schema, the volume of the data is already so large that the Blazegraph triplestore which is used in the Wikidata Query Service reaches its capacity limit, 
preventing any integration \cite{wdqssearchteam2022wdqs} (see Sec.~2).

In this paper, we introduce \textit{SemOpenAlex}, an extremely large RDF dataset of the academic landscape with its publications, authors, sources, institutions, concepts, and publishers. SemOpenAlex consists of more than 26 billion semantic triples and includes over 249 million publications from all academic disciplines. It is based on our rich ontology (see Sec.~\ref{subchapter_SOA_ontology}) and includes links to other LOD sources such as Wikidata, Wikipedia, and the MAKG. To ensure easy and efficient use of SemOpenAlex's integration with the LOD cloud, we provide a public SPARQL endpoint. In addition, we provide a sophisticated semantic search interface that allows users to retrieve real-time information about contained entities and their semantic relationships (e.g., displaying co-authors or an author's top concepts -- information, which is not directly contained in the database but obtained through semantic reasoning). We also provide the full RDF data snapshots to enable big data analysis. Due to the large size of SemOpenAlex and the ever-increasing number of scientific publications being integrated into SemOpenAlex, we have established a pipeline using AWS for regularly updating SemOpenAlex entirely without any service interruptions. 
Additionally, to use SemOpenAlex in downstream applications, we trained state-of-the-art knowledge graph entity embeddings. 
By reusing existing ontologies whenever possible, we ensure system interoperability in accordance with FAIR principles \cite{wilkinson_fair_2016} and pave the way for the integration of SemOpenAlex into the Linked Open Data Cloud. We fill the gap left by the discontinuation of MAKG by providing monthly updates that facilitate ongoing monitoring of an author's scientific impact, tracking of award-winning research, and other use cases using our data \cite{huang_towards_2022,wagner_nobel_2015}. By making SemOpenAlex free and unrestricted, we empower research communities across all disciplines to use the data it contains and integrate it into their projects. Initial use cases and production systems that use SemOpenAlex already exist (see Sec.~\ref{chapter_use-cases}).

Overall, we make the following contributions: 
\begin{enumerate}
    \item We create an \textit{ontology} for SemOpenAlex reusing common vocabularies.
    \item We create the SemOpenAlex \textit{knowledge graph} in RDF, covering 26 billion triples, 
    and provide all \textit{SemOpenAlex} data, code, and services for public access at \mbox{\url{https://semopenalex.org/}}:
     \begin{enumerate}
        \item We provide monthly updated RDF data snapshots free of charge %
        on AWS S3 at \url{s3://semopenalex} (via browser: \url{https://semopenalex.s3.amazonaws.com/browse.html}), 
        accepted as AWS Open Data project.\footnote{The AWS Open Data Sponsorship program covers the cost of storing and retrieving all SemOpenAlex data, ensuring the long-term sustainability of our project. 
        Upon request, it was confirmed that Zenodo does not support the provision of SemOpenAlex data due to its size.} 
        \item We make all URIs of SemOpenAlex resolvable, allowing SemOpenAlex to be part of the Linked Open Data cloud.\footnotemark\footnotetext{See, e.g. \texttt{curl -H "Accept:text/n3" https://semopenalex.org/work/W4239696231}}
        \item We index all data in a triple store and make it publicly available via a SPARQL endpoint (\url{https://semopenalex.org/sparql}).
        \item We provide a semantic search interface including entity disambiguation to access, search, and visualize the knowledge graph and its statistical key figures in real time. 
    \end{enumerate}

    \item %
    We provide state-of-the-art knowledge graph embeddings for the entities represented in SemOpenAlex using high-performance computing.

\end{enumerate}

In the following, we first discuss related work (see Sec.~\ref{chapter_related_work}) 
and describe the 
SemOpenAlex 
ontology and RDF data 
(see Sec.~\ref{chapter_soa}), 
before presenting the SemOpenAlex entity embeddings (see Sec.~\ref{chapter_embeddings}). 
Subsequently, we outline existing and potential use cases 
(see Sec.~\ref{chapter_use-cases}), before we conclude the paper (see Sec.~\ref{chapter_conclusions}).

\section{Related Work}
\label{chapter_related_work}

A comparison of scholarly RDF datasets is presented in Table~\ref{tab:comparison}. It is obvious from the table that SemOpenAlex (1) is the only RDF KG that follows the Linked Data Principles, (2) is fully open, (3) contains a vast amount of bibliographic information from all scientific disciplines, and (4) is regularly updated, making it a valuable resource 
in various contexts (see Sec.~\ref{chapter_use-cases}). 

\setlength{\tabcolsep}{4pt}
\begin{table}[tb]
 \centering
 \caption{Statistical comparison of scholarly RDF datasets.\tablefootnote{OpenAIRE as of March 2021, AceKG as of 2018, Wikidata as of Dec. 2022,  the OpenCitations Index of Crossref open DOI-to-DOI citations (COCI) as of Oct. 2022, the MAKG as of March 2021, and SemOpenAlex as of March 2023}}
 \label{tab:comparison}
\resizebox{\textwidth}{!}{
\begin{tabular}{lrrrrrr}
\toprule
&        OpenAIRE & AceKG   & Wikidata     & COCI      & MAKG   & SemOpenAlex \\
 \midrule
\# Works   &  145M  &  62M   &  42M      & 76M    & 239M     & 249M  \\
\# Triples &  1.4B  & 3.13B  & -     & 1.4B   & 8B      & 26.4B \\
\# References &  0  &  480M &    288M      & 1.4B   & 1.4B   & 1.7B      \\
Snapshot size & 100GB & 113GB &  120GB   & 1.5TB &    1.4TB & 1.7TB \\
\midrule
Regular updates &  &  & (\checkmark)  & \checkmark &     & \checkmark \\
\midrule
SPARQL endpoint &  &  & \checkmark  & \checkmark & \checkmark & \checkmark \\
\midrule
Entity embeddings 
&  &       &       &          &\checkmark& \checkmark \\   
\bottomrule
\end{tabular}
}
\end{table}

The OpenAIRE Research Graph provides open and free access to metadata of 
145 million 
publications, datasets, and software via an API, a SPARQL endpoint, and 
database dumps \cite{manghi_openaire_2022}. However, not only is the number of publications significantly lower than in SemOpenAlex but on May 8, 2023, OpenAIRE 
stopped its 
LOD services and closed the 
SPARQL endpoint.\footnote{See \url{https://www.openaire.eu/pausing-our-lod-services}.}

WikiCite\footnote{See \url{http://wikicite.org/}.} has incorporated bibliographic metadata into Wikidata, but SemOpenAlex covers considerably more metadata (e.g., 249M papers vs. 42M), 
including additional properties such as papers' abstracts. While using Wikidata as a central KG and regularly importing SemOpenAlex information seems logical, the scalability of 
the Blazegraph triplestore backend which hosts the   
Wikidata Query Service is limited, and Wikimedia has announced a plan to delete scholarly articles in case of 
bulk imports.\footnote{See \url{https://m.wikidata.org/wiki/Wikidata:SPARQL_query_service/WDQS_backend_update/Blazegraph_failure_playbook}.}

AceKG~\cite{wang_acekg_2018} is a database containing 62 million publications, along with academic details related to authors, fields of study, venues, and institutes. AceKG data is modeled in RDF. However, unlike our approach, it does not use existing vocabularies, lacks a publicly available triple store, and does not offer continuous updates. All data is sourced from a company's database. %

OpenCitations focuses on publications and their citation relationships \cite{peroni_setting_2015}. Specifically, it covers metadata about publications and their citations, but not descriptions of affiliated organizations (institutions) or hosting conferences and journals (venues). OpenCitations includes several datasets, including the OpenCitations Index of Crossref Open DOI-to-DOI Citations (COCI) with 76 million items to date, and smaller datasets such as the OpenCitations Corpus (OCC) and OpenCitations in Context Corpus (CCC) \cite{peroni_opencitations_2020}.

The Microsoft Academic Knowledge Graph (MAKG) is 
based on the Microsoft Academic Graph (MAG), containing information on publications, authors, institutions, venues, and concepts \cite{farber_makg_2019,sinha_overview_2015}. The MAKG has high coverage across scientific domains and has enabled novel use cases. However, it will no longer be updated 
due to lack of source data \cite{microsoft_research_next_2021}. Several analyses have assessed the MAG and MAKG, revealing the need for improvements in areas such as citation accuracy, concept assignment, and disambiguation \cite{herrmannova_analysis_2016,visser_large-scale_2021,chen_glimpse_2020,wang_microsoft_2020}. 
Compared to MAKG, SemOpenAlex provides a similar schema, but provides fresh data that is in addition cleaned by an author name disambiguation provided by OpenAlex and a neater mapping of concepts to papers using the Simple Knowledge Organization System (SKOS) ontology \cite{farber_microsoft_2022,tay_goodbye_2021}.

Further notable scholarly KGs are the DBLP KG\footnote{See \url{https://www.dagstuhl.de/en/institute/news/2022/dblp-in-rdf}} and the Open Research Knowledge Graph
(ORKG) \cite{ORKG2020}. DBLP provides only high-quality metadata about computer science publications, resulting in a coverage of roughly 6 million publications \cite{aleman-meza_swetodblp_2007}. ORKG is a project that aims to provide a KG infrastructure for semantically capturing and representing the content of research papers \cite{auer_towards_2018,jaradeh_open_2019}. ORKG contains a relatively small set of more than 25,000 publications, however, with 
many RDF statements, indicating considerable semantic richness. Due to their different focuses, SemOpenAlex can complement ORKG as an LOD data source: while SemOpenAlex provides a broad basis of metadata about a massive amount of publications and related entities in RDF (with a focus on high coverage, see Table \ref{tab:distribution-entity-types}), ORKG focuses on modeling scientific contributions as well as methodology aspects, which are manually curated (with a focus on high data quality and key insights of papers).

\section{SemOpenAlex}
\label{chapter_soa}

In the following, we describe the design of the SemOpenAlex ontology (Sec. 3.1) and the process of generating SemOpenAlex data (Sec. 3.2). We also explain how we publish and enable user interaction with the data (Sec. 3.3), and present key statistics of the KG (Sec. 3.4). Furthermore, we evaluate to what extent SemOpenAlex meets linked data set descriptions and rankings (Sec. 3.5).

\begin{table}[tb]
    \centering
    \caption{SemOpenAlex entity types and number of instances (as of March 2023).}
    \label{tab:distribution-entity-types}
    \begin{small}
    \begin{tabular}{lr}
    \toprule
     Entity Type    &  \# Instances \\
    \midrule
         Work    & 249,450,604 \\
         Author    &  135,360,159 \\
         Source    & 226,413  \\
         Institution    & 108,618  \\
         Concept    & 65,073  \\
         Publisher    & 7,017  \\
    \bottomrule
    \end{tabular}
    \end{small}
\end{table}

\subsection{Ontology of SemOpenAlex}
\label{subchapter_SOA_ontology}

\begin{sidewaysfigure}
\centering
\includegraphics[width=\textwidth]{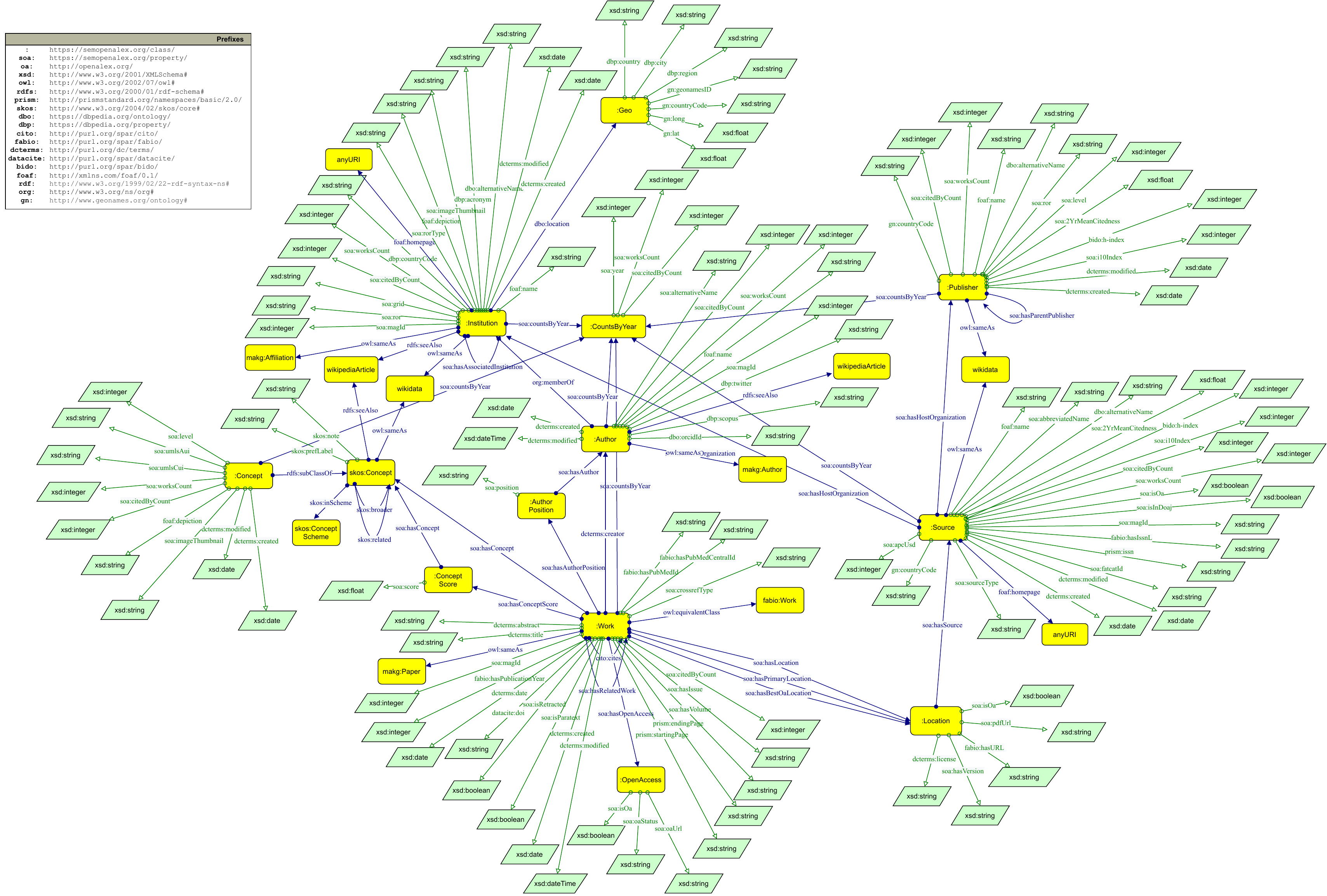}
\caption{Ontology of SemOpenAlex. }
\label{fig:semopenalex-schema}
\end{sidewaysfigure}

We developed an ontology following the best practices of ontology engineering reusing as much existing vocabulary as possible. An overview of the entity types, the object properties, and the data type properties is provided in Fig.~\ref{fig:semopenalex-schema}. Overall, the ontology of SemOpenAlex covers \textit{13 entity types}, including the main entity types \textit{works, authors, institutions, sources, publishers and concepts}, as well as \textit{87 relation types}.

We reused the vocabularies listed in Table~\ref{tab:prefixes}. To describe publications, researchers, and institutions, we leveraged established Semantic Publishing and Referencing (SPAR) ontologies \cite{vrandecic_spar_2018}, such as FaBiO and CiTO. FaBiO is used to describe specific identifiers such as a work’s PubMedID, while CiTO represents citing relationships between works. For bibliographic metadata, such as a work’s publication date and abstract, we used the Dublin Core ontology (DCterms). To represent more generic features and relations, we relied on cross-domain ontologies such as DBpedia and the W3 Organization Ontology (W3 ORG). The works are classified using a concept hierarchy, which we represented in a SKOS vocabulary of 65k SKOS concepts and semantic relations (\texttt{skos:broader} and \texttt{skos:related}). The concepts are further linked with Wikidata entities, allowing for additional interoperability and providing multi-lingual labels.

\begin{table}[tb]
    \centering
        \caption{Used ontologies, their corresponding prefixes and namespace.}
    \label{tab:prefixes}
    \begin{footnotesize}
    \begin{tabular}{lll}
    \toprule
    Ontology & Prefix & Associated URI \\
    \midrule
\texttt{SemOpenAlex} & \texttt{:}     & \url{https://semopenalex.org/class/}\\
\texttt{SemOpenAlex} & \texttt{soa:}  & \url{https://semopenalex.org/property/}\\
\texttt{OpenAlex}    & \texttt{oa:}   & \url{http://openalex.org/}\\
\texttt{XML Schema}  & \texttt{xsd:}  & \url{http://www.w3.org /2001/XMLSchema\#}\\
\texttt{OWL}         & \texttt{owl:}  & \url{http://www.w3.org/2002/07/owl\#}\\
\texttt{RDF}         & \texttt{rdf:}  & \url{http://www.w3.org/1999/02/22-rdf-syntax-ns\#}\\
\texttt{RDF Schema}  & \texttt{rdfs:}  & \url{http://www.w3.org/2000/01/rdf-schema\#}\\
\texttt{Dubin Core}  & \texttt{dcterms:}  & \url{http://purl.org/dc/terms/}\\
\texttt{CiTO}        & \texttt{cito:}  & \url{http://purl.org/spar/cito/}\\
\texttt{FaBiO}       & \texttt{fabio:}  & \url{http://purl.org/spar/fabio/}\\
\texttt{BiDO}       & \texttt{bido:}  & \url{http://purl.org/spar/bido/}\\
\texttt{DataCite}    & \texttt{datacite:} & \url{http://purl.org/spar/datacite} \\
\texttt{PRISM}       & \texttt{prism:}  & \url{http://prismstandard.org/namespaces/basic/2.0/}\\
\texttt{DBpedia}     & \texttt{dbo:}  & \url{https://dbpedia.org/ontology/}\\
\texttt{DBpedia}     & \texttt{dbp:}  & \url{https://dbpedia.org/property/}\\
\texttt{FOAF}        & \texttt{foaf:}  & \url{http://xmlns.com/foaf/0.1/}\\
\texttt{W3 ORG}      & \texttt{org:}  & \url{http://www.w3.org/ns/org\#}\\
\texttt{GeoNames}    & \texttt{gn:}  & \url{https://www.geonames.org/ontology\#}\\
\texttt{SKOS}        & \texttt{skos:} & \url{http://www.w3.org/2004/02/skos/core\#}\\
\bottomrule
    \end{tabular}
    \end{footnotesize}
\end{table}

\subsection{Knowledge Graph Creation Process}
\label{subchapter_SOA_transformation_process}

The raw OpenAlex data was presumably designed for data processing (e.g., abstracts are provided as inverted index and not provided as one string).
To create an RDF KG based on the OpenAlex dump files, major changes in the data formatting and the data modeling are necessary. In the following, we outline the essential steps of this transformation process.

\subsubsection*{Transformation. }
We carry out a number of distinct steps for the transformation that can be reproduced via the code in our GitHub repository.\footnote{See \url{https://github.com/metaphacts/semopenalex}.}

\begin{enumerate}
    \item \textit{Data Preprocessing:} 
    We download the OpenAlex snapshot in compressed \texttt{.jsonl} format from its AWS S3 bucket and use the Python multiprocessing package 
    for efficient parallel processing of the large amount of data. To ensure valid triple generation according to the \textit{W3C RDF 1.1 Concepts and Abstract Syntax}\footnote{See \url{https://www.w3.org/TR/rdf11-concepts/\#section-Graph-Literal}.} later, we remove problematic characters from literal values, such as non-escaped backslashes in URLs or newlines in publication titles. Additionally, we convert the abstracts, which are included in OpenAlex as an inverted index, to plain text to improve accessibility.

    \item \textit{RDF Generation:} We transform the preprocessed data from JSON into RDF according to the ontology shown in Fig. \ref{fig:semopenalex-schema}.
    For the generation of the triples, we draw on the rdflib Python package,\footnote{See \url{https://github.com/RDFLib/rdflib/}.} which offers functionality to handle, process and validate RDF data. 
    During triple serialization, we create a buffer subgraph that is written once a fixed number of statements is reached to reduce the number of I/O operations. 
    In total, we generate 26,401,183,867 RDF triples given the data snapshot as of 2023-03-28.
    
    \item \textit{Compression and Deployment:} 
    The RDF data generated for SemOpenAlex takes up 1.7TB in the TriG format\footnote{TriG is an extension of Turtle, extended to support representing a complete RDF dataset (see \url{https://www.w3.org/TR/trig/}).} when uncompressed. To make the data more manageable, we compress it into .gz archives, resulting in a reduction of over 80\% in file size to 232GB. These compressed files are then imported into the GraphDB triple store and made available for download as an open snapshot. Additionally, we provide a data sample on GitHub.

\end{enumerate}

\subsubsection*{Update Mechanism}
To ensure that SemOpenAlex remains up-to-date, we perform the transformation process described earlier on a monthly basis, which involves downloading the latest OpenAlex snapshot. This enables us to observe temporal dynamics in the data, and ensures that SemOpenAlex provides the most recent information available. The updated version of the data is available through all three access points (RDF dump, SPARQL endpoint, and visual interface). The update process is semi-automated and takes approximately five days to complete on an external server instance. We use one AWS instance to provide SemOpenAlex services and one instance to process the next SemOpenAlex release. Changes to SemOpenAlex data resulting from changes in the raw OpenAlex files are tracked using announcements via the OpenAlex mailing list. Several adaptations have been performed in this way in the past.

\subsection{Data Publishing and User Interaction}
\label{subchapter_SOA_deployment}

\begin{figure}[t]
    \centering
    \includegraphics[width=1.0\linewidth]{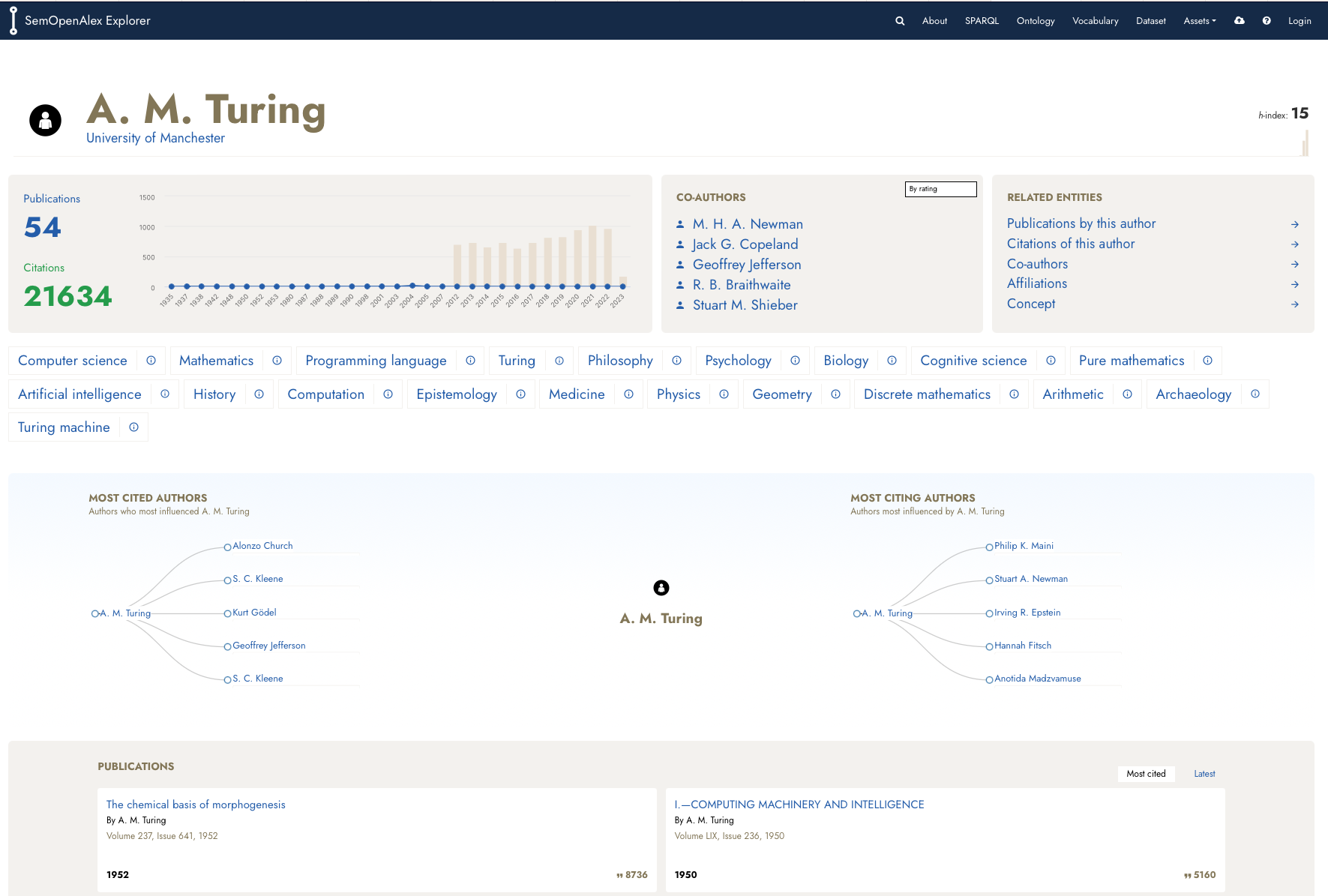}
    \caption{Author overview page for A.M. Turing, accessible at \url{https://semopenalex.org/author/A2430569270}.}
    \label{fig:metaphactory_turing_overview}
\end{figure}

Our KG is publicly accessible 
at 
\url{https://semopenalex.org/}. We utilize the metaphactory knowledge graph platform \cite{DBLP:journals/semweb/HaaseHKNT19} on top of a GraphDB triple store to deploy the KG. metaphactory serves as a Linked Data publication platform and ensures that the URIs of SemOpenAlex are fully resolvable. The data is published in machine-readable RDF formats as well as human-readable HTML-based templates using content negotiation. Fig.~\ref{fig:metaphactory_turing_overview} displays the page for the URI \url{https://semopenalex.org/author/A2430569270}.

Among other features, the interface provided for SemOpenAlex enables users to: (1)~access SemOpenAlex through a search interface with filtering options; (2)~visualize arbitrarily large sub-graphs for objects and relations of interest; (3)~formulate and execute SPARQL queries to retrieve objects from the graph using a provided SPARQL endpoint; (4)~examine the ontology of SemOpenAlex; (5)~obtain key statistics for each object in SemOpenAlex in a dashboard, as shown in the screenshot in Fig.~\ref{fig:metaphactory_turing_overview}; (6)~assess the underlying multi-level concept 
hierarchy; and (7)~interact with further linked entities such as co-authors or concepts and access external resources such as links to Wikidata.

\subsection{Key Statistics of SemOpenAlex
and Example SPARQL Queries}
\label{subchapter_SOA_summary}

\begin{figure}[tb]
    \centering
    \includegraphics[width=0.75\linewidth]{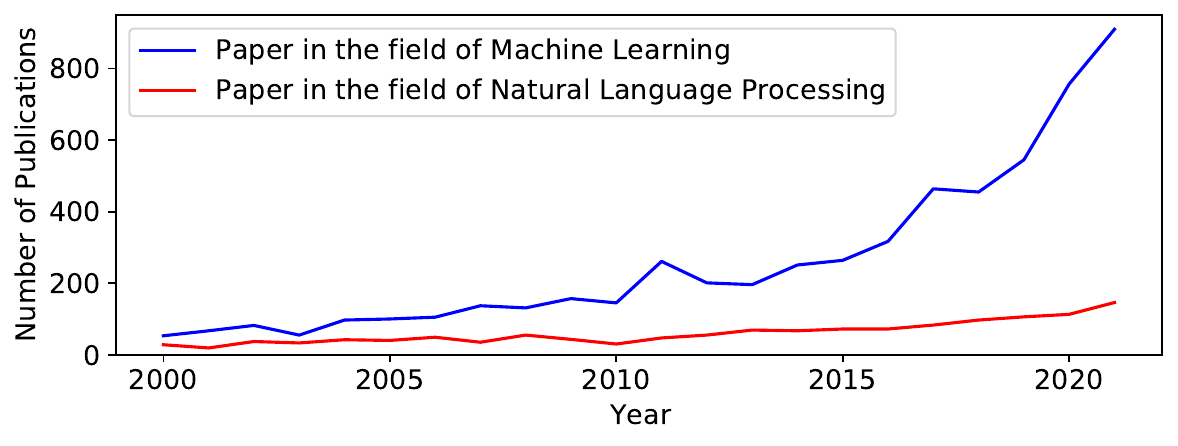}
    \caption{Number of publications published in 
    machine learning and natural language processing by researchers from Karlsruhe Institute of Technology.}    \label{fig:concepts-kit}
\end{figure}

\begin{table}[tbp]
\centering
   \begin{minipage}{0.35\textwidth}
    \centering
    \begin{small}
        \caption{Number of institution for the countries with the most institutions.}  
        \label{tab:numer-institutions-countries}
    \begin{tabular}{lr}
    \toprule
     Country    &  \# Institutions \\
    \midrule
         US    &  32,814 \\
         GB    & 7,743 \\
              DE    & 5,096  \\
              CN    & 4,856  \\
              JP    & 4,031  \\
              FR    & 3,965  \\
              IN    & 3,731  \\
              CA    & 3,498  \\
    \bottomrule
    \end{tabular}
    \end{small}
  \end{minipage}
  ~
  \begin{minipage}{0.57\textwidth}
      \centering
      \begin{figure}[H]
       \includegraphics[width=\textwidth]{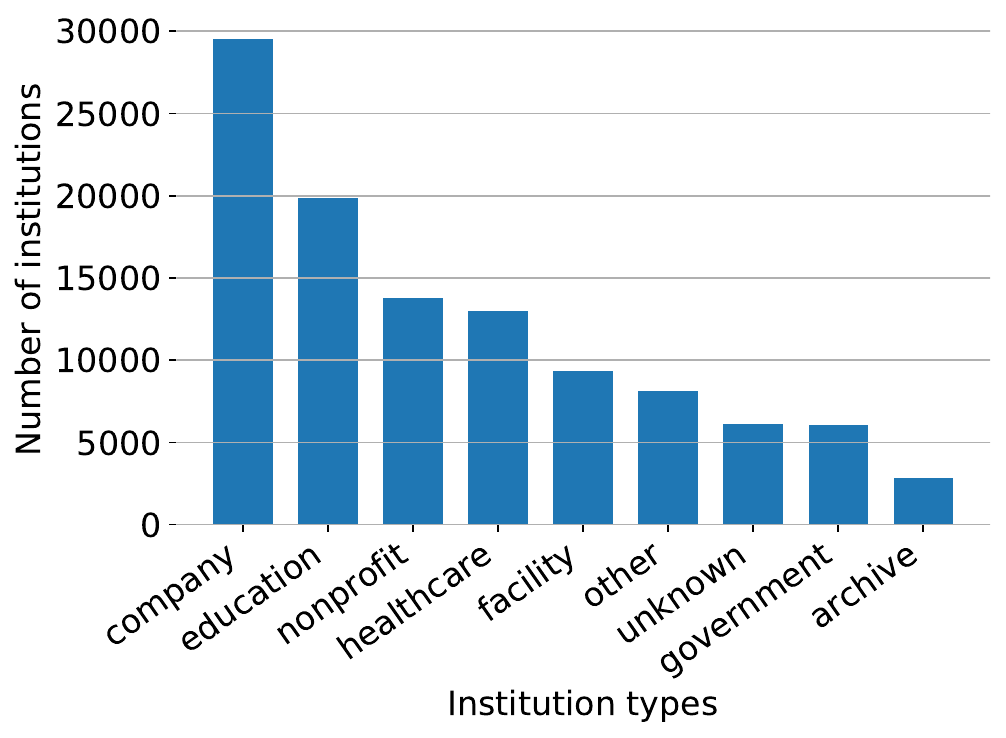}
       \caption{Distribution of institution types.} %
       \label{fig:distribution-institution-types}
      \end{figure}
  \end{minipage}
\end{table}

In this subsection, we present several statistics that we generated based on queries using our SPARQL endpoint. We provide the queries on GitHub.

Fig. \ref{fig:concepts-kit} shows the number of papers published in the field of machine learning and natural language processing by researchers from Karlsruhe Institute of Technology from 2000 to 2021. While the number of machine learning papers received a sharp increase from 2015, the number of papers in the field of natural language processing increased at a rather constant rate. SemOpenAlex enables institutions to create such relevant key figures and trends in the context of strategic controlling in a simple and cost-free way.

SemOpenAlex covers the worldwide scientific landscape and contains publications from institutions around the globe. In total, institutions from 225 different countries are included. The 8 countries with the highest number of institutions are shown in Table \ref{tab:numer-institutions-countries}.

By distinguishing between eight types of institutions, SemOpenAlex enables differentiated data analyses. In Fig.~\ref{fig:distribution-institution-types}, we can see the distribution of the 108,618 unique institutions across the different types. We can see that the majority of the organizations are companies, followed by educational and nonprofit institutions.

List.~\ref{lst:query1} shows an example of how SemOpenAlex can be queried with SPARQL. 
This query retrieves the top 100 most cited papers in the field of semantic web, along with their citation counts and first authors. It is worth noting that this query cannot be executed on other scholarly KGs like the MAKG, as they do not cover information about the author's position for a given paper.

\begin{lstlisting}[captionpos=b,label={lst:query1},caption={Querying the top 100 most cited papers with the concept ``Semantic Web'' as well as their citation count and first author.},
   basicstyle=\scriptsize\ttfamily,frame=single,breaklines=true,float=t] 
PREFIX foaf: <http://xmlns.com/foaf/0.1/>
PREFIX xsd: <http://www.w3.org/2001/XMLSchema#>
PREFIX dcterms: <http://purl.org/dc/terms/>
PREFIX soa: <https://semopenalex.org/property/>
PREFIX skos: <http://www.w3.org/2004/02/skos/core#>

SELECT DISTINCT ?paperTitle ?citedByCount ?firstAuthorName 
WHERE {
  ?paper dcterms:title ?paperTitle .
  ?paper soa:hasConcept ?Concept .
  ?Concept skos:prefLabel "Semantic Web"^^xsd:string .
  ?paper soa:citedByCount ?citedByCount .
  ?paper soa:hasAuthorPosition ?authorPosition .
  ?authorPosition soa:position "first"^^xsd:string .
  ?authorPosition soa:hasAuthor ?firstAuthor .
  ?firstAuthor foaf:name ?firstAuthorName .
} 
ORDER BY DESC(?citedByCount)
LIMIT 100
\end{lstlisting}

\subsection{Linked Data Set Descriptions and Ratings}
\label{subchapter_description_rating}

Following the licensing model of the underlying OpenAlex data,\footnote{See \url{https://openalex.org/about}.} we provide all SemOpenAlex data under the \textit{CC0 license}, which grants users the right to freely build upon, enhance, and reuse the works for any purpose without restriction, paving the way for other researchers and software engineers to build upon SemOpenAlex in any context. 
The RDF data files are available for unrestricted and free download as they are hosted with the AWS Open Data program.\footnote{See \url{https://aws.amazon.com/opendata/open-data-sponsorship-program/}.}

We can categorize SemOpenAlex according to the two kinds of 5-star rating schemes in the Linked Data context:
\begin{itemize}
   \item \textit{Tim Berners-Lee's 5-star deployment scheme for Open Data}\footnote{See \url{http://5stardata.info/}.}:
   Our SemOpenAlex RDF dataset is a 5-star data set according to this scheme, because we provide our data in RDF (leading to 4 stars) and the (1)~entity URIs are linked to Wikidata, Wikipedia and the MAKG and (2)~our vocabulary URIs to other vocabularies (leading to 5 stars).
  \item \textit{Linked Data vocabulary star rating} \cite{Janowicz2014}: 
    This rating is intended to rate the use of vocabulary within Linked (Open) Data. 
    By providing a turtle file, by linking our vocabulary to other vocabularies (see the SPAR ontologies), we are able to provide the vocabulary with 4 stars.
\end{itemize}

\newpage{}
Aside from the SemOpenAlex RDF documents, we provide the following linked data set descriptions (all available at \url{https://semopenalex.org/}):
\vspace{-0.2cm}
\begin{itemize}
\item \textit{Turtle}: We provide our ontology as a Turtle file describing the used classes, object properties, and data type properties.
\item \textit{VoID}: We provide a VoID file to describe our linked data set with an RDF schema vocabulary.
\end{itemize}
\vspace{-0.4cm}

\section{Graph Embeddings for SemOpenAlex}
\label{chapter_embeddings}

Apart from creating and providing the SemOpenAlex data set and services (e.g., the SPARQL endpoint), we computed embeddings for all SemOpenAlex entities. 
Entity embeddings have proven to be useful as implicit knowledge representations in a variety of scenarios, as we describe in Sec.~\ref{chapter_use-cases}. 
Based on the SemOpenAlex data in RDF, we trained entity embeddings based on several state-of-the-art embedding techniques and compared the performance of the respective results with regard to link prediction tasks. 
Specifically, we applied 
the following approaches: 
TransE \cite{bordes_translating_2013}, DistMult \cite{yang_distmult_2014}, ComplEx \cite{trouillon_complex_2016}, a GraphSAGE neural network \cite{hamilton_inductive_2018}, and a graph attention network \cite{velickovic_graph_2017}. %
To address the nontrivial challenges associated with training on SemOpenAlex as a very large knowledge graph, we employed the Marius framework \cite{waleffe_mariusgnn_2022}. Marius\footnote{See \url{https://marius-project.org}.} is designed to optimize resource utilization by pipelining hard disk, CPU, and GPU memory during training, thereby reducing idle times. 
In our evaluation, we opted for a configuration of 100 embedding dimensions, a batch size of 16,000, and trained for 3 epochs on a high-performance computing system (bwUniCluster~2.0) using Python 3.7, Marius 0.0.2, PyTorch 1.9.1, and CUDA 11.2.2. 
These parameters are in line with previous research on large-scale entity embeddings 
\cite{farber_microsoft_2022}. 

The computational effort required for the different embedding techniques varied, with the GraphSAGE and the graph attention network approaches requiring the most memory. These methods used up to 716GB of CPU RAM and took the longest time to train, with each epoch taking roughly 24 hours. 
Despite the resource-intensive nature of the GraphSAGE and graph attention network approaches, 
DistMult yielded the highest mean reciprocal rank (MRR) score in our link prediction evaluation (see all evaluation results on GitHub). Therefore, we provide the DistMult-based embedding vectors for all entities online.\footnote{See \url{https://doi.org/10.5281/zenodo.7912776}.}

\section{Use Cases of SemOpenAlex}
\label{chapter_use-cases}

Scholarly KGs have proven to be a valuable data basis for various use cases and scenarios, such as analyzing research dynamics between academia and industry \cite{angioni_aida_2021}, scientific impact quantification \cite{schindler_investigating_2020,huang_towards_2022}, and linking research data sets to publications \cite{farber_data_2021}. 
This is also reflected in the high number of citations of the 
reference publications of the MAG~\cite{Sinha2015} and MAKG~\cite{farber_makg_2019}.\footnote{Sinha et al.\cite{Sinha2015} have obtained 1,041 citations and Färber \cite{farber_makg_2019} has obtained 115 citations as of April 28, 2023, according to Google Scholar.} 
In the following, we focus on existing and potential use cases of SemOpenAlex.

\textbf{Scholarly Big Data Analytics and Large-Scale Scientific Impact Quantification.}
SemOpenAlex can serve for scientific impact quantification and innovation management. For instance, OpenAlex has been utilized as a comprehensive and reliable data source to rank researchers and institutions worldwide on \texttt{research.com}.\footnote{See \url{https://research.com/university/materials-science/humboldt-university-of-berlin}.} InnoGraph is a new project that leverages OpenAlex to represent innovation ecosystems as a KG for innovation management and forecasting \cite{massri2023towards}. By using SemOpenAlex as underlying database for such projects and efforts, the need to deal with cumbersome data integration issues can be reduced. 
Currently, universities such as KIT rely on paid scholarly services like those from Springer Nature for measuring their performance and ranking as a university \cite{marginson2014university}. However, in the future, these institutions can use SemOpenAlex as a free database to run analytics and evaluations on all relevant publications and associated entities.

\textbf{Scholarly Search and Recommender Systems.} 
Recommendation systems --  both content-based and collaborative filtering-based -- have become increasingly important in academia to help scientists navigate the overwhelming amount of available information resulting from the exponential increase in the number of publications. 
In this paper, we provide entity embeddings for nearly all existing entities in the scientific landscape, which can be used directly to build state-of-the-art recommender systems. These systems can recommend items such as papers to read and cite, as well as venues and collaborators \cite{hu2020heterogeneous}. SemOpenAlex can be utilized to make these recommendations explainable, as symbolic information from the KG can be shown to the user.
Due to SemOpenAlex's rich ontology, including various entity types, SemOpenAlex can serve as a realistic dataset for training and evaluating state-of-the-art graph neural networks designed for heterogeneous information networks and with a specific focus on scalability and semantics. 
Moreover, our rich KG can be utilized to provide recommendations in complex scenarios, such as finding the optimal consortium for large, possibly interdisciplinary research projects.
In the context of semantic search, SemOpenAlex can be used 
for 
entity linking, annotating scientific texts \cite{Faerber2019JCDL} or tables \cite{lou2023s2abel} 
for enhanced search capabilities.

\textbf{Semantic Scientific Publishing. }
SemOpenAlex is a part of the Linked Open Data Cloud and contains links to other data sources such as Wikidata, Wikipedia, and MAKG. As a result, it significantly contributes to the use of linked data in areas such as digital libraries and information retrieval \cite{Carrasco2016}. SemOpenAlex has a unique selling point among available scientific knowledge graphs, with its coverage of publications worldwide and across all scientific disciplines, totaling around 250 million publications (see Table~\ref{tab:distribution-entity-types}), and its regular updates.
SemOpenAlex can serve as a central catalog for publications, researchers, and research artifacts, to which other data repositories and KGs can link. This creates an opportunity to use SemOpenAlex as a basis for modeling scientific artifacts, such as datasets, scientific methods, and AI models, and thus beyond SemOpenAlex' current scope. This information may be modeled in separate, interlinked KGs or as part of SemOpenAlex in the future. 
For instance, the Data Set Knowledge Graph \cite{farber_data_2021}, which currently links 600,000 publications in which datasets are mentioned to the MAKG, can now link datasets to papers in SemOpenAlex. Similarly, semantic representations of datasets and scientific methods \cite{DBLP:conf/aaai/FarberAS21}, as well as 
representations of scientific facts and claims mentioned in full-text articles \cite{Fathalla2017}, can be linked to publications and authors in SemOpenAlex to provide rich context information as explanations of academic recommender systems. 
Furthermore, links between SemOpenAlex and KGs modeling AI models and their energy consumption, such as the Green AI Knowledge Graph \cite{farber2022green}, can be used to combine previously isolated data for performing complex analytics. In this way, questions of strategic controlling, such as ``How green are the AI models developed at my institution?'' \cite{farber2022green}, can be automatically answered. 
Finally, it makes sense to link full-text paper collections to SemOpenAlex, for instance, to leverage its concept schema, since SemOpenAlex applies concept tags to all its papers published globally and across all scientific fields. An excellent example of an existing paper collection linked to SemOpenAlex is unarXive 2022 \cite{saier2023unarxive}, sourced from two million arXiv papers.

\textbf{Research Project Management and Modeling.} %
KGs have become increasingly important in supporting research projects by providing a structured representation of various research entities and their relationships \cite{DBLP:conf/semweb/DiefenbachWA21}. These project-specific KGs encapsulate a diverse range of research entities, such as topics, methods, tasks, materials, organizations, researchers, their skills, interests, and activities, as well as research outputs and project outcomes.
To facilitate the development and support of KGs for research projects, SemOpenAlex serves as a knowledge hub by providing existing data on project participants and relevant research. Researchers can use tools and vocabularies provided by the Competency Management Ontology \cite{DBLP:conf/semweb/Heist021} to seamlessly describe their skills, current research interests, and activities in terms of the entities already contained in SemOpenAlex. Moreover, SemOpenAlex's concept hierarchy allows for the construction of ontologies for specific research domains, streamlining research tasks such as performing a state-of-the-art analysis for a research area. Existing resources from SemOpenAlex can be integrated into KG-based project bibliographies, enhancing collaboration between researchers through resource sharing.

SemOpenAlex has already been used 
to provide a comprehensive and structured overview of research projects. In particular, personalized dashboards have been created by metaphacts that display 
recently added publications from SemOpenAlex that are relevant to the current research context. Newly created resources within a project, such as research papers and datasets, can also be described and linked to SemOpenAlex. Ultimately, published results become a valuable part of SemOpenAlex.

\textbf{Groundwork for Scientific Publishing in the Future.} 
One can envision that the working style of researchers will considerably change in the next few decades \cite{Hoffman2018,Jaradeh2019}. For instance, publications might not be published in PDF format any more, but in either an annotated version of it (with information about the claims, the used methods, the data sets, the evaluation results, and so on) or in the form of a flexible publication form,  in which authors can change the content and, in particular, citations, over time. SemOpenAlex can be easily combined with new such data sets due to its structure in RDF. Furthermore, ORKG is an ongoing effort that targets the semantic representation of papers and their scientific contributions. We argue that SemOpenAlex can be used as data basis for ORKG in the sense that with SemOpenAlex, users do not need to take care of first creating papers and authors in the ORKG, but to directly import or link the corresponding information from SemOpenAlex, which has its focus on being a comprehensive KG covering all scientific publications worldwide. 

\textbf{Knowledge-Guided Language Models.} 
Large language models, including ChatGPT and GPT-4, have been criticized for their lack of explainability and their failure to provide reliable in-text citations to reference literature. Often, when citations are provided, they are incorrect and reflect ``hallucinations''. 
In this context, SemOpenAlex represents a valuable repository for guiding language models in providing reliable references to scientific literature and as a basis for text-editing generative models. With metadata of 250 million scientific works, SemOpenAlex can serve as a valuable resource for source attribution and improving the accuracy and quality of scientific writing generated by these models. 

\textbf{Benchmarking.} 
SemOpenAlex is a prime example of big data, fulfilling the ``4 V's'' criteria: it is very large, 
with a wide variety of information types (including papers, authors, institutions, venues, and various data formats), contains uncertainties, and is updated periodically. This makes it suitable for benchmarking systems and approaches, particularly in the context of querying large, realistic KGs \cite{DBLP:conf/edbt/Cossu0L18}. In fact, the MAKG has already been used for this purpose \cite{bassani2022multi}
and we expect SemOpenAlex to follow suit.

\section{Conclusions}
\label{chapter_conclusions}

In this paper, we presented a comprehensive RDF dataset with over 26 billion triples covering scholarly data across all scientific disciplines. We outlined the creation process of this dataset 
and discussed its characteristics. Our dataset supports complex analyses through SPARQL querying. By making the SPARQL endpoint publicly available and the URIs resolvable, we enriched the Linked Open Data cloud with a valuable source of information in the field of academic publishing. We offer RDF dumps, linked dataset descriptions, a SPARQL endpoint, and trained entity embeddings online at \url{https://semopenalex.org/}. In the future, we plan to incorporate metadata about funding programs to enable in-depth and comprehensive evaluations of funding lines of governments and institutions \cite{dziezyc2022effectiveness,jonkers2016research,DBLP:conf/semweb/DiefenbachWA21}.

\noindent
\textbf{Acknowledgments.} 
This work was partially supported by the German Federal Ministry of Education and Research (BMBF) as part of the project IIDI (01IS21026D). The authors acknowledge support by the state of Baden-Württemberg through bwHPC.

\bibliographystyle{splncs}
\bibliography{bibliography}

\end{document}